\documentclass[12pt]{article}

\usepackage{amssymb}
\usepackage{amsmath}

\usepackage{epsfig}

\begin{document}

\title{\bf Interactions of cosmic neutrinos with nucleons in the RS model}

\author{A.V. Kisselev\thanks{Email address: alexandre.kisselev@ihep.ru} \\ \\
\em  Institute for High Energy Physics, 142281 Protvino,
Russia}

\date{}

\maketitle

\thispagestyle{empty}

\bigskip

\begin{abstract}
We consider the scattering of the brane fields due to $t$-channel
massive graviton exchanges in the Randall-Sundrum model. The
eikonal amplitude is analytically calculated and both differential
and total neutrino-nucleon cross sections are estimated. The event
rate of quasi-horizontal air showers induced by cosmic neutrinos,
which can be detected at the Pierre Auger Observatory, is
presented for two different fluxes of cosmogenic neutrinos.
\end{abstract}


\section{Introduction}

The detection of air showers induced by ultra-high energy
neutrinos may help to solve many important problems, such as
propagation of cosmic neutrinos to the Earth and their
interactions with the nucleons at energies around tens (hundreds)
of TeV. In this energy region, neutrino-nucleon interactions may
be strong due to a new physics. There is a large class of models
in a space-time with extra spacial dimensions which result in a
new TeV phenomenology. In the present paper we will consider an
approach with non-factorizable metric proposed in
Refs.~\cite{Randall:99,Randall:99*}, and study a scattering of the
SM fields in this scenario.

The RS model~\cite{Randall:99,Randall:99*} is a model of gravity
in a slice of a 5-dimensional Anti-de-Sitter space (AdS$_5$) with
a single extra dimension compactified to the orbifold $S^1/Z_2$.
The metric is of the form:
\begin{equation}\label{01}
ds^2 = e^{-2 \kappa |y|} \, \eta_{\mu \nu} \, dx^{\mu} \, dx^{\nu}
+ dy^2.
\end{equation}
Here $y = r_c \theta$ ($0 \leqslant \theta \leqslant \pi$), $r_c$
is a "radius" of extra dimension, and parameter $\kappa$ defines
the scalar (negative) curvature of the space.

From a 4-dimensional action one can derive the relation:
\begin{equation}\label{02}
\bar{M}_{Pl}^2 = \frac{M^3}{\kappa} \left( 1 - e^{-2 \pi \kappa r}
\right) \simeq \frac{M^3}{\kappa},
\end{equation}
which means that $\kappa \sim \bar{M}_{Pl} \sim M$,  with $M$
being a 5-dimensional Planck scale.

We will consider so-called RS1 model~\cite{Randall:99} which has
two 3-dimensional branes with equal but opposite in sign tensions
which are located at the point $y = \pi r_c$ (called the TeV
brane) and at the point $y = 0$ (referred to as the Planck brane).
All SM fields are constrained to the TeV brane, while the gravity
propagates in the bulk (all spacial dimensions).

From the point of view of an observer located on the TeV brane,
there exists an infinite number of graviton Kaluza-Klein (KK)
excitations with masses:
\begin{equation}\label{04}
m_n = x_n \kappa \, e^{-\pi \kappa r_c}, \qquad n=1,2 \ldots,
\end{equation}
where $x_n$ are zeros of the Bessel function $J_1(x)$:%
\footnote{The first four values of $x_n$ are 3.83, 7.02, 10.17,
and 13.32.}
\begin{equation}\label{05}
J_1(x_n) = 0, \qquad n=1,2 \ldots
\end{equation}

By using a linear expansion of the metric, one can derive the
interaction Lagrangian:
\begin{equation}\label{06}
\mathcal{L} = - \frac{1}{\bar{M}_{Pl}} \, T_{\mu \nu} \,
h^{(0)}_{\mu \nu} - \frac{1}{\Lambda_{\pi}} \, T_{\mu \nu} \,
\sum_{n=1}^{\infty} h^{(n)}_{\mu \nu},
\end{equation}
where
\begin{equation}\label{08}
\Lambda_{\pi} = \bar{M}_{Pl} \, e^{-\pi \kappa r_c}
\end{equation}
is a physical scale on the TeV brane. It can be chosen as small as
1 TeV  for a thick slice of the AdS$_5$, $r_c \simeq 12/\kappa
\simeq 60 \, l_{Pl}$. We see from \eqref{06} that couplings of all
massive states are only suppressed by $\Lambda_{\pi}^{-1}$, while
the zero mode couples with usual strength defined by the reduced
Planck mass $\bar{M}_{Pl} = M_{Pl}/\sqrt{8 \pi}$.

\begin{figure}[htpb]
\centering
\epsfig{figure=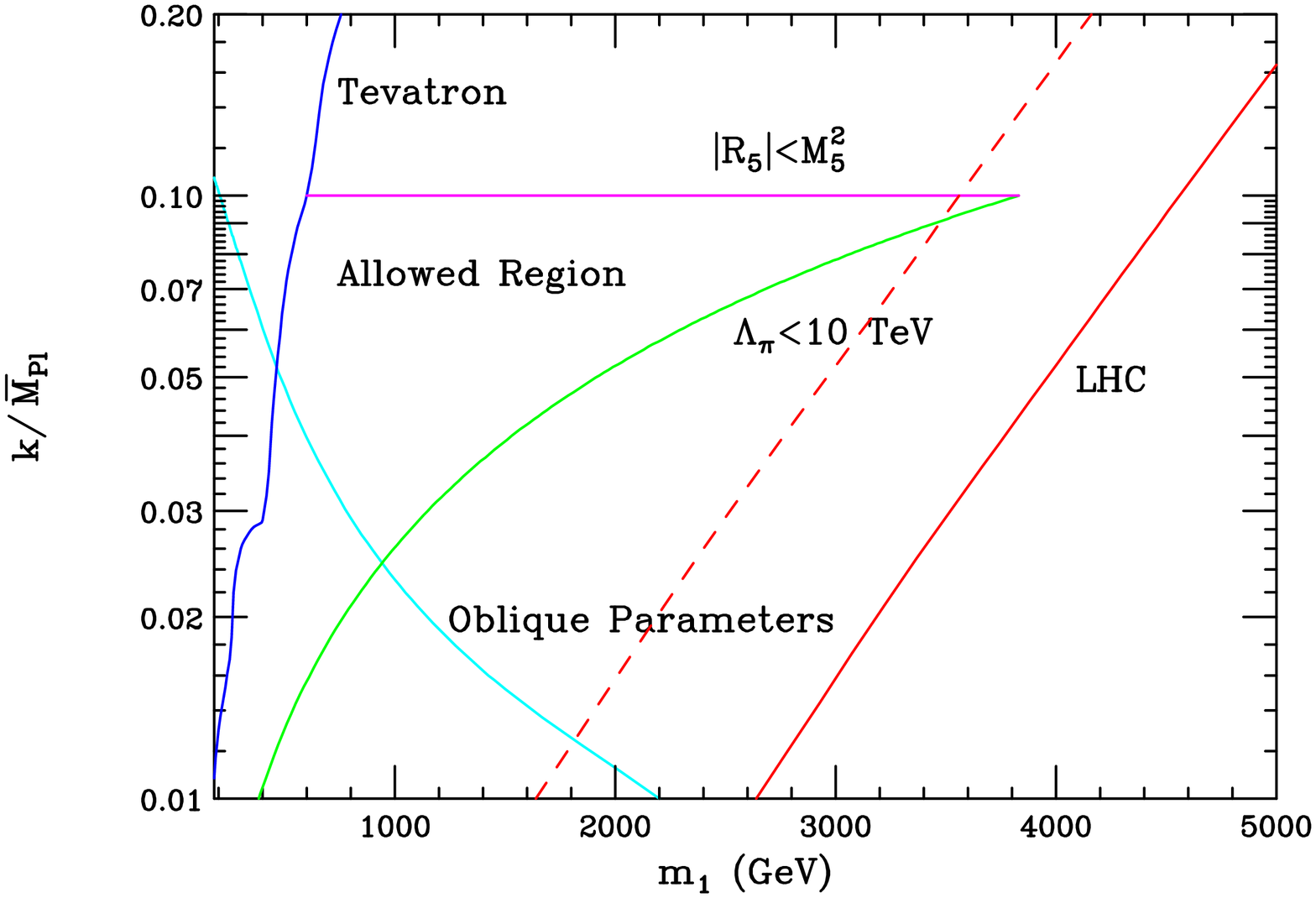,height=14cm,width=8cm,angle=90}
\caption{Experimental and theoretical constrains on the RS model
in the two-parameter plane $\kappa/\bar{M}_{Pl}$ --
$m_1$~\cite{Davoudiasl:01}. The allowed region lies in the center
as indicated.}
\label{fig:bounds}
\end{figure}

The main phenomenological parameters of the model are the scale
$\Lambda_{\pi}$ and the ratio
\begin{equation}\label{10}
\mu = \frac{\kappa}{\bar{M}_{Pl}}.
\end{equation}
The present experimental data together with theoretical bounds on
the curvature of the AdS$_5$ restrict an allowed region for the
variable $\mu$ (see, for instance, Fig.~\ref{fig:bounds} taken
from Ref.~\cite{Davoudiasl:01}):
\begin{equation}\label{11}
0.01 \lesssim \mu \lesssim 0.1.
\end{equation}
The allowed value of $\mu$ is restricted by so-called naturalness
and requiring a 5-dimensional curvature to be small enough to
consider a linearized gravity on the brane. Thus, the lightest
masses of the KK graviton modes, $m_n = x_n \mu \Lambda_{\pi}$,
are of the order of 1 TeV.

Our paper has the following structure. In the next Section we
consider interactions of the SM fields on the brane in the RS
model induced by an exchange of massive gravitons. The eikonal
amplitude is calculated and an elastic cross section for different
sets of the RS parameters and/or invariant energy is estimated. In
Sec.~3 we use these results to study the scattering of ultra-high
energy cosmic neutrino off the atmospheric nucleons. The
neutrino-nucleon cross section is calculated and an event rate of
quasi-horizontal neutrino events expected at the Auger Observatory
is presented. Our conclusions and discussions is a topic of
Sect.~4.

\section{Eikonal amplitude in the RS model}

In what follows, we will employ the zero width approximation for
the graviton KK resonances. The Born amplitude corresponding to
$t$-channel exchange looks like (both the massless mode and KK
gravitons contribute):
\begin{equation}\label{12}
A_B(s,t) =  \frac{8 \pi G_N s^2}{-t} + \frac{s^2}{\Lambda_{\pi}^2}
\sum_{n=1}^{\infty} \frac{1}{-t + m_n^2}.
\end{equation}
The sum in \eqref{12} converges very rapidly in $n$, since $x_n =
\pi \, (n + 1/4) + \mathrm{O(n^{-1})}$~\cite{Watson}. We consider
the scattering of two particles \emph{living on the TeV brane}.
Thus, in Eq.~\eqref{12} $t$ means a 4-dimensional momentum
transfer which is well-defined and conserved.

Let us underline that in \eqref{12} we sum spin-two particles with
different KK numbers $n$ (non-reggeized KK gravitons). In more
general approach, one should sum Regge trajectories $\alpha_n(t)$
which are numerated by $n$ (KK-charged gravireggeons). For the ADD
model, it was done in Refs.~\cite{Kisselev:04}. The results of
Refs.~\cite{Giudice:02,Kisselev:04} can be reproduced in the limit
$\alpha_n(t) \rightarrow 2$. As for the RS model, results on a
gravireggeon contribution to the eikonal amplitude will be
presented in a forthcoming paper~\cite{Kisselev:05}.

Generally, the massive KK states may decay to a pair of SM
particles. The partial widths are proportional to
$m_n^3/\Lambda_{\pi}^2$, where $m_n$ is the mass of the resonance.
In particular, the partial decay widths to massless gauge bosons,
fermions, and a pair of Higgs are%
\footnote{These expressions can be obtained by a replacement
$\bar{M}_{Pl}^{-2} \rightarrow \Lambda_{\pi}^{-2}$ in
corresponding formulae derived for large extra dimensions in
Ref.~\cite{Han:99}. We have also neglected masses of the SM
particles, since $m_{_{SM}} \ll m_n$.}
\begin{eqnarray}\label{13}
\Gamma (h^{(n)} \rightarrow VV) &=& N_V \, a \, \frac{m_n^3}{40
\pi \Lambda_{\pi}^2},
\nonumber \\
\Gamma (h^{(n)} \rightarrow f \bar{f}) &=& N_f \, \frac{m_n^3}{160
\pi \Lambda_{\pi}^2},
\nonumber \\
\Gamma (h^{(n)} \rightarrow H \bar{H}) &=& \frac{m_n^3}{480 \pi
\Lambda_{\pi}^2}.
\end{eqnarray}
Here $N_V = 1(8)$ for photons and electroweak bosons (gluons),
$N_f = 1(3)$ for a lepton (quark pair) mode, and $a=1/2$ for
identical particles. Then for the total width of the massive KK
graviton in the RS model, $\Gamma_n$, we get the estimate (see
 also \cite{Dvergsnes:03}):
\begin{equation}\label{14}
\Gamma_n \simeq m_n \, (0.31 \, \mu \, x_n)^2.
\end{equation}
Since the sum which we are interested in converges very rapidly in
$n$ (see a comment after Eq.~\eqref{12}), we conclude from
\eqref{14} and \eqref{11} that effectively $\Gamma_n/2 \ll m_n$.

The sum~\eqref{12} can be calculated analytically by the use of
the formula~\cite{Watson}:
\begin{equation}\label{15}
\sum_{n=1}^{\infty} \frac{1}{ z_{n, \nu}^2 - z^2} =
\frac{J_{\nu + 1}(z)}{2 \, z\,J_{\nu}(z)},
\end{equation}
where $z_{n, \, \nu}$ ($n=1,2 \ldots$) are zeros of the function
$z^{-\nu} J_{\nu}(z)$. As a result, we obtain:
\begin{equation}\label{16}
A_B(s,t) =  \frac{8 \pi G_N s^2}{-t} + \frac{s^2}{2 \mu
\Lambda_{\pi}^3} \, \frac{1}{\sqrt{-t}} \, \frac{I_2(v)}{I_1(v)}
\end{equation}
Here $I_i(z)$ ($i=1,2$) are modified Bessel functions and $v =
\sqrt{-t}/\mu \Lambda_{\pi}$. Taking into account properties of
$I_i(x)$, we conclude from \eqref{16} that a contribution of the
massive graviton modes dominates at large $|t|$:
\begin{equation}\label{17}
A_B(s,t) \Big|_{|t| \gg \mu \Lambda_{\pi}} \simeq \frac{s^2}{2 \mu
\Lambda_{\pi}^3} \, \frac{1}{\sqrt{|t|}}.
\end{equation}
Note, we would get another asymptotics in $t$, namely, $A_B(s,t)
\sim |t|^{-1}$, if we sum only finite number of the massive
gravitons.

As it was shown in Ref.~\cite{Giudice:02}, it is ladder diagrams
that makes a leading contribution of the KK gravitons to an
amplitude and results in the eikonal representation for the
amplitude ($q^2 = -t$):
\begin{equation}\label{18}
A^{eik}(s,t) = 2is \! \int d^2 b \, e^{i q \, b} \left[ 1 - e^{i
\chi(s,b)} \right],
\end{equation}
with the eikonal given by
\begin{equation}\label{19}
\chi(s, b) = \frac{1}{4\pi s} \int dq \, q \, J_0(q \, b) \,
A_B(s, -q^2).
\end{equation}

The proper accounting for the massless mode has been presented in
Ref.~\cite{Kisselev:03}. The result is the following:
\begin{eqnarray}\label{20}
&& A^{eik}(s,t)  = e^{i \phi_4} \, \Bigg\{ \frac{8\pi G_N s}{-t}
\, \frac{\Gamma(1 - i G_N s)}{\Gamma(1 + i G_N s)}
\nonumber \\
&&  + \, 4\pi i s \, (-t)^{-i G_N s} \! \int\limits_0^{\infty} \,
db \, b^{1 -2i G_N s} \, J_0 \left( b \sqrt{|t|} \right) \Big[ 1 -
e^{i \chi_{mass} (s,b)} \Big]  \Bigg\}.
\end{eqnarray}
Here $\chi_{mass}(s,b)$ denotes a contribution the massive modes
to the eikonal, $G_N$ is the Newton constant, and $\phi_4$ is a
4-dimensional (infinite) phase. The first term in the RHS of
Eq.~\eqref{20} is well-known 4-dimensional result derived by
different methods in Refs.~\cite{Hooft:87}. It is negligible at
any conceivable energy and momentum transfer, and we can write (up
to a phase factor):
\begin{equation}\label{21}
A^{eik}(s,t) \simeq  4\pi i s \! \int\limits_0^{\infty} \, db \, b
\, J_0 \left( b \sqrt{|t|} \right) \Big[ 1 - e^{i \chi_{mass}
(s,b)} \Big].
\end{equation}

It follows from \eqref{16} that the eikonal depends on two
dimensionless variables, $s/\Lambda_{\pi}^2$, and
\begin{equation}\label{22}
u = b \mu \Lambda_{\pi},
\end{equation}
and it looks like
\begin{equation}\label{24}
\tilde{\chi}_{mass}(s,u) \equiv \chi_{mass}(s,\frac{u}{\mu
\Lambda_{\pi}}) = \frac{1}{8\pi} \, \frac{s}{\Lambda_{\pi}^2} \!
\int\limits_0^{\infty} \, dv \, J_0(u v) \, \frac{I_2(v)}{I_1(v)}.
\end{equation}
The eikonal~\eqref{24} is very well approximated by the following
expression (see Appendix for details):
\begin{equation}\label{25}
\tilde{\chi}_{mass}(s,u) \simeq \frac{\sqrt{3}}{16\pi} \,
\frac{s}{u \Lambda_{\pi}^2} \, \exp(-2 \sqrt{3} \, u).
\end{equation}

At $\sqrt{s}  \gg 5 \, \Lambda_{\pi}$, the eikonal is
exponentially small outside the region
\begin{equation}\label{26}
b \lesssim b_0(s) = \frac{1}{\sqrt{3} \, \mu \Lambda_{\pi}} \, \ln
\frac{\sqrt{s}}{\Lambda_{\pi}}.
\end{equation}
At $b \rightarrow 0$, it is proportional to $b^{-1}$. Thus, we can
roughly estimate the high-energy behavior of elastic cross
section:
\begin{equation}\label{28}
\sigma_{el}(s)  \simeq \frac{\pi}{3\,(\mu \Lambda_{\pi})^2} \,
\ln^2 \frac{\sqrt{s}}{\Lambda_{\pi}} \thickapprox
\frac{\pi}{m_1^2} \, \ln^2 \frac{s}{\Lambda_{\pi}^2},
\end{equation}
where $m_1$ is a mass of a lightest KK graviton.

Let us underline that the Froissart-Martin like formula \eqref{28}
describes the contribution of the massive graviton modes. The
presence of the \emph{massless} graviton in the theory should
result in \emph{infinite} elastic and total cross
sections~\cite{Petrov:02}. However, its contribution can be safely
neglected in our further calculations.

We can rewrite Eq.~\eqref{21} in the form
\begin{equation}\label{30}
A^{eik}(s,t) \simeq  4\pi i \, \frac{s}{(\mu \Lambda_{\pi})^2} \!
\int\limits_0^{\infty} \, du \, u \, J_0 \left( u
\frac{\sqrt{-t}}{\mu \Lambda_{\pi}} \right) \Big[ 1 - e^{i
\tilde{\chi}_{mass} (s,u)} \Big].
\end{equation}
Correspondingly, the differential cross section in dimensionless
variable
\begin{equation}\label{32}
y = \frac{-t}{s}
\end{equation}
is defined by
\begin{equation}\label{34}
\frac{d \sigma_{el}}{dy} = \frac{1}{16 \pi s} \,
|A^{eik}(s,-ys)|^2,
\end{equation}
and we get the estimate
\begin{equation}\label{36}
\frac{d \sigma_{el}}{dy} \Big|_{y=0} \simeq \frac{\pi \, s}{36
\,(\mu \Lambda_{\pi})^4} \, \ln^4 \frac{\sqrt{s}}{\Lambda_{\pi}}.
\end{equation}

It follows from \eqref{34}, \eqref{30} that $d\sigma_{el}/dy$
depends only on variable $y$~\eqref{32}, parameter $\mu$, and the
ratio $\sqrt{s}/\Lambda_{\pi}$, in addition to the dimensional
factor $(\mu \Lambda_{\pi})^{-2}$ which defines a magnitude of the
cross section. In particular, we have $d\sigma_{el}/dy |_{y=0} = s
\, (\mu \Lambda_{\pi})^{-4} \, f(s/\Lambda_{\pi}^2)$, and
$\sigma_{el} = (\mu \Lambda_{\pi})^{-2} \, g(s/\Lambda_{\pi}^2, \,
\mu)$, were $f(x)$ and $g(x,y)$ are dimensionless functions
defined via the eikonal.

\begin{figure}
\centering
\epsfig{figure=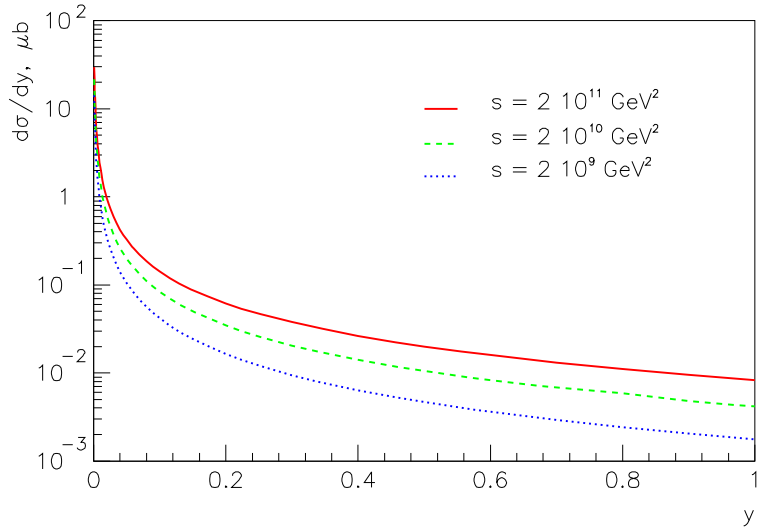,height=7cm,width=10cm}
\caption{The differential cross section as a function of
dimensionless variable $y$ for three fixed values of the invariant
energy. The parameters of the RS model are chosen to be
$\Lambda_{\pi}=$ 2 TeV, and $\mu=0.05$.}
\label{fig:dsigma elas}
\end{figure}

\begin{figure}
\centering
\epsfig{figure=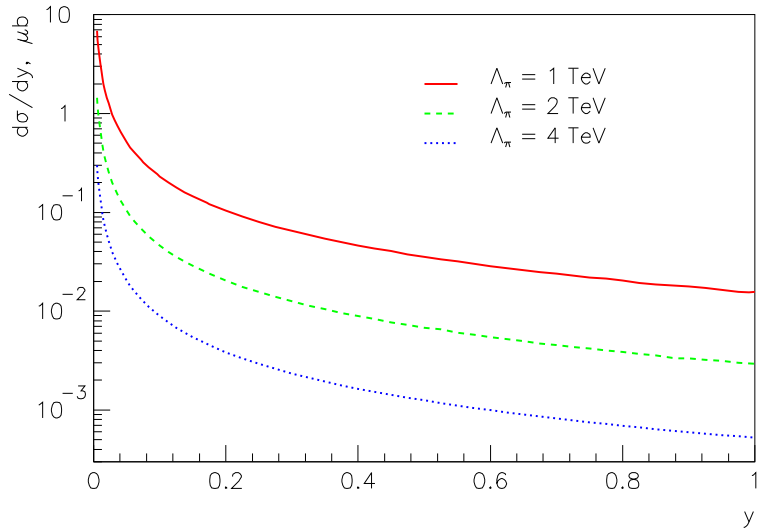,height=7cm,width=10cm}
\caption{The differential cross section as a function of variable
$y$ for three values of the mass scale $\Lambda_{\pi}$ at fixed
energy $s=2 \cdot 10^{11}$ GeV$^2$ (with the parameter
$\mu=0.1$).}
\label{fig:Lambda dependence}
\end{figure}

\begin{figure}
\centering
\epsfig{figure=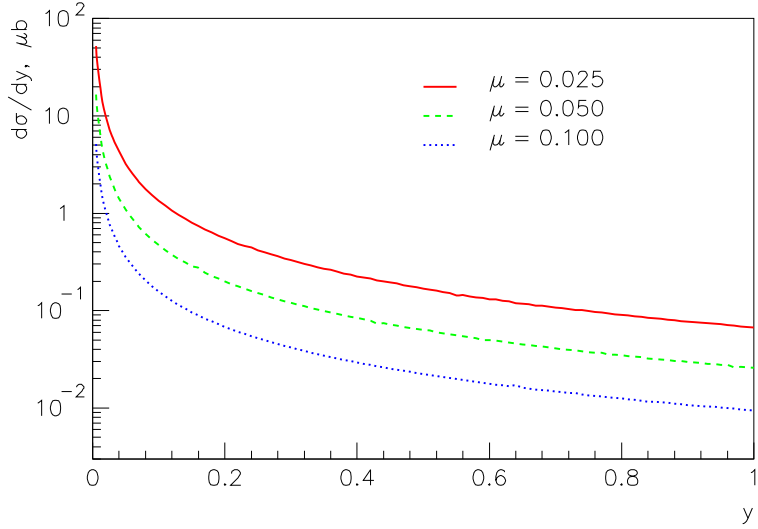,height=7cm,width=10cm}
\caption{The differential cross section as a function of variable
$y$ for three values of the RS parameter $\mu$ at fixed energy
$s=2 \cdot 10^{10}$ GeV$^2$ (with the scale $\Lambda_{\pi} = $ 1
TeV).}
\label{fig:mu dependence}
\end{figure}

\begin{figure}
\centering
\epsfig{figure=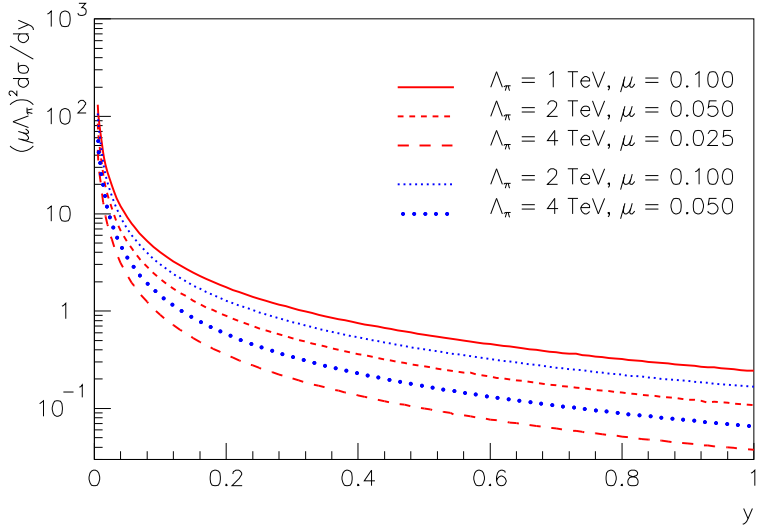,height=7cm,width=10cm}
\caption{The reduced (dimensionless) differential cross section as
a function of variable $y$ for different parameter sets
($\Lambda_{\pi}$, $\mu$) at invariant energy $s=2 \cdot 10^{10}$
GeV$^2$. The product $\mu \Lambda_{\pi}$ is taken to be 100 GeV
(200 GeV) for three first (two last) sets.}
\label{fig:lambda-mu total dependence}
\end{figure}

The results of our calculations with the use of formulae
\eqref{30} \eqref{34}, and \eqref{24} are presented in
Fig.~\ref{fig:dsigma elas}-\ref{fig:lambda-mu total dependence}.
The curves in Fig.~\ref{fig:dsigma elas} which show an energy
dependence of the cross section were obtained for $\Lambda_{\pi}=$
2 TeV, and $\mu=0.05$. The dependence of the differential cross
section on the parameter $\Lambda_{\pi}$ at $s=2 \cdot 10^{11}$
GeV$^2$, $\mu = 0.1$ is presented in Fig.~\ref{fig:Lambda
dependence}. Next Fig.~\ref{fig:mu dependence} demonstrates the
dependence of $d\sigma/dy$ on the parameter $\mu$ at $s=2 \cdot
10^{10}$ GeV$^2$, $\Lambda_{\pi} = $ 1 TeV. Finally, in
Fig.~\ref{fig:lambda-mu total dependence} the reduced differential
cross section (namely, multiplied by the factor $(\mu
\Lambda_{\pi})^2$) is shown for several sets ($\Lambda_{\pi},
\mu$).

\section{Neutrino-nucleon cross section and neutrino induced
air showers}

Let us now estimate a neutrino-nucleon differential cross section
as a function of variable $y$. The neutrino scatters off quarks
and gluons which are distributed inside the nucleon. Thus,
neutrino-nucleon cross section is presented by
\begin{equation}\label{38}
\frac{d\sigma^{\nu N}(s)}{dy} = \int\limits_{x_{min}}^1 dx \sum_i
f_i(x,M^2) \, \frac{d\sigma^{\nu i}(\hat{s})}{dy},
\end{equation}
where $f_i(x,M^2)$ is a distribution of parton $i$ in momentum
fraction $x$, and $\hat{s} = xs$ is an invariant energy of a
partonic subprocess. The partonic differential cross section,
$d\sigma^{\nu i}(\hat{s})/dy$, is defined via the eikonal
\eqref{25} taken at the energy $\sqrt{\hat{s}}$.

We use a set of parton distribution functions (PDFs) from
Ref.~\cite{Alekhin:02} based on an analysis of existing deep
inelastic data in the next-to-leading order QCD approximation in
the fixed-flavor-number scheme. The extraction of the PDFs is
performed in \cite{Alekhin:02} simultaneously with the value of
the strong coupling and high-twist contributions to structure
functions. The PDFs are available in the region $10^{-7} < x < 1$,
$2.5 \text{ GeV}^2 < Q^2 < 5.6 \cdot 10^7 \text{
GeV}^2$~\cite{Alekhin:02}. So, no extrapolation in variable $x$ is
needed.

We put $x_{min} = \Lambda_{\pi}^2/s$ in \eqref{36}. Since the
eikonal is effectively cut at $b = b_0(\hat{s})$~\eqref{26}, we
take the mass scale in PDFs to be $M = 1/b_0(\sqrt{\hat{s}})$. The
effective impact parameter $b_0$ is much smaller than the size of
the nucleon. Thus, our assumption that the neutrino interacts with
the constituents of the nucleon and probes its inner structure, is
well justified.

The differential cross section as a function of $y$, the energy
fraction deposited from the neutrino to the nucleon, is presented
in Fig.~\ref{fig:dsigma_10EeV} for the neutrino energy $E_{\nu} =
10$ EeV and three sets of parameters of the RS model.

\begin{figure}
\centering
\epsfig{figure=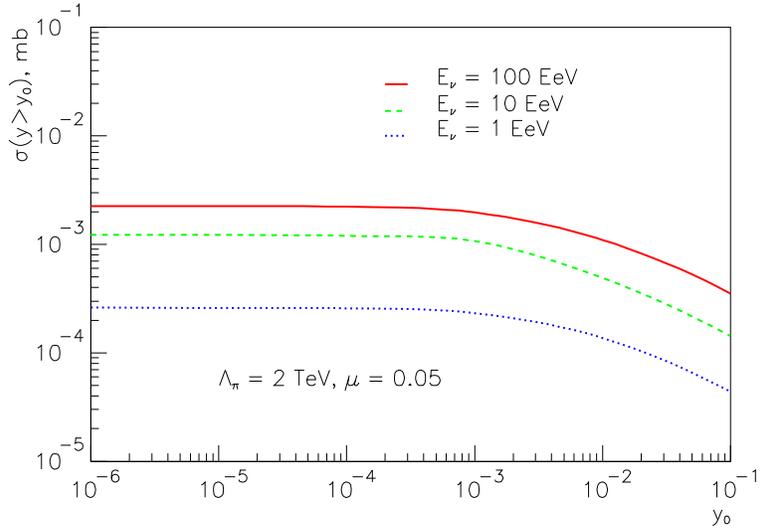,height=7cm,width=10cm}
\caption{The inelastic neutrino-proton cross-section obtained by
integrating the differential cross section in the region $y_0
\leqslant y \leqslant 1$ as a function of $y_0$, the minimal
fraction of the neutrino energy deposited to the proton.}
\label{fig:sigma_y0_1}
\end{figure}

\begin{figure}
\centering
\epsfig{figure=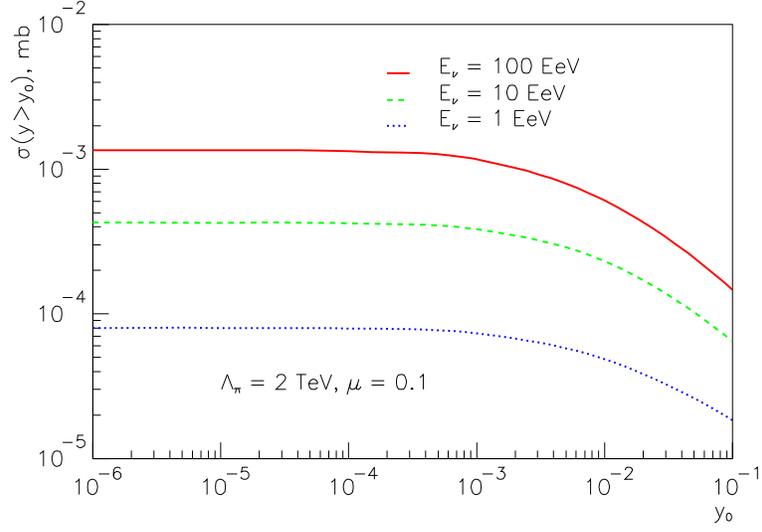,height=7cm,width=10cm}
\caption{The same as in Fig.~\ref{fig:sigma_y0_1} but for the
different value of $\mu=0.1$.}
\label{fig:sigma_y0_2}
\end{figure}

\begin{figure}
\centering
\epsfig{figure=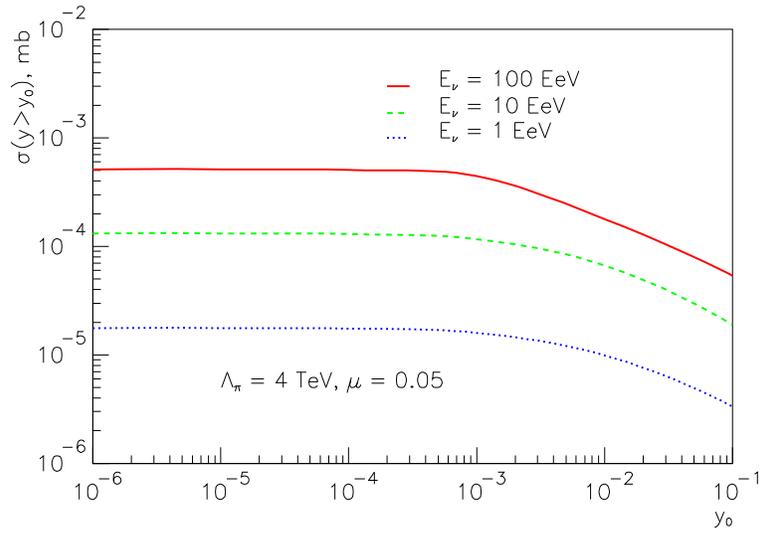,height=7cm,width=10cm}
\caption{The same as in Fig.~\ref{fig:sigma_y0_1} but for the
different value of $\Lambda_{\pi}=$ 4 TeV.}
\label{fig:sigma_y0_3}
\end{figure}

In order to estimate an effective range of variable $y$ which
contributes to the neutrino-nucleon cross section, we have
calculated a quantity
\begin{equation}\label{40}
\sigma^{\nu N} (y > y_0) = \int\limits_{y_0}^1 \frac{d \sigma^{\nu
N}}{dy},
\end{equation}
where $y_0$ is a minimum fraction of energy lost by the neutrino
(deposited to the nucleon). The dependence of the quantity
$\sigma_{el} (y > y_0)$ on $y_0$ at different values of the
neutrino energy $E_{\nu}$ is shown in Fig.~\ref{fig:sigma_y0_1}
for $\Lambda_{\pi} = 2$ TeV, $\mu = 0.05$. Next two figures,
Fig.~\ref{fig:sigma_y0_2} and Fig.~~\ref{fig:sigma_y0_3}, show the
dependence of $\sigma_{el} (y
> y_0)$ on the parameters $\mu$ and $\Lambda_{\pi}$.

Ultra-high energy cosmic neutrinos have not yet been detected (see
non-observation of neutrino-induced events reported by the Fly's
Eye~\cite{Fly's Eye}, the AGASA~\cite{AGASA} and the
RICE~\cite{RICE} collaborations). A number of experiments under
construction will allow to measure fluxes of such neutrinos within
the next few years. Among them are Pierre Auger Observatory,
IceCube neutrino telescope at the South Pole, Anita radio detector
for a balloon flights around the South Pole, as well as EUSO,
SalSA and OWL proposals. We will consider the first
possibility~\cite{Auger}.

The number of horizontal hadronic air showers with the energy
$E_{sh}$ larger than a threshold energy $E_{th}$, initiated by
neutrino-nucleon interactions, is given by
\begin{eqnarray}\label{42}
N_{sh} &=& T N_A \int\limits_{E_{th}}^{E_{max}} dE \, \Big[
\sum_{i = e, \, \mu, \, \tau} \! \Phi_{\nu_i}(E) \int\limits_0^1
dy \,\, \frac{d\sigma_{\nu N}^{grav} (E)}{dy} \, {\cal A}(y E) \,
\theta(yE - E_{th})
\nonumber \\
&+&  \sum_{i = e, \, \mu, \, \tau} \! \Phi_{\nu_i}(E) \,
\sigma_{\nu_i N}^{SM}(E) \, {\cal A}(\bar{y}_i E_{\nu_i}) \,
\theta(\bar{y}_i E - E_{th}) \Big],
\end{eqnarray}
where $N_A = 6.022 \cdot 10^{23} \text{ g}^{-1}$, $T$ is a time
interval (one year, in our case), and ${\cal A}(E)$ is a detector
acceptance as a function of a shower energy (in units of km$^3$
steradian water equivalent $= 10^{15}$ g). The quantity
$\Phi_{\nu_i}(E)$ in \eqref{42} is a flux of the neutrino of type
$i$. Both neutrino and antineutrino are assumed in the sums in
Eq.~\eqref{42}. The product $E \, \Phi_{\nu_i}(E)$ is in units of
cm$^{-2}$ yr$^{-1}$. We have taken into account that the energy of
the shower resulted from the gravitational interaction is equal to
$yE$, and that this interaction is universal for all types of
neutrinos.

For the energy distribution of the neutrino in the SM processes,
we have used the approximation $d\sigma_{\nu_i N}^{SM}(y,E)/dy
\simeq \sigma (E) \, \delta(y - \bar{y_i})$. The inelasticity
$\bar{y}_i$ defines a mean fraction of the neutrino energy
deposited into the shower in a corresponding SM process. We have
put $\bar{y}_e =1$ for SM \emph{charged} current interactions
initiated by electronic neutrino,  while for SM \emph{neutral}
interactions initiated by $\nu_e$ and for $\nu_{\mu}/
\nu_{\tau}$-events we have taken $\bar{y}_e = \bar{y}_{\mu} =
\bar{y}_{\tau} =0.2$~\cite{Sigl:98}.

The number of extensive quasi-horizontal showers induced by
so-called cosmogenic neutrinos which can be detected by the array
of the southern site of the Pierre Auger Observatory, is presented
in Table~\ref{tab:cosmogenic flux} for several sets of the RS
parameters. These values of parameters are chosen in such a way in
order not to violate experimental and theoretical bounds presented
in Fig.~\ref{fig:bounds}.%
\footnote{Remember that in our notations $\kappa/\bar{M}_{Pl}
\equiv \mu$, $m_1 \simeq 3.83 \, \mu \Lambda_{\pi}$.}
The cosmogenic neutrino flux is taken from
Refs.~\cite{Protheroe:96}, assuming $E_{max} = 3 \cdot 10^{21}$
eV. The acceptance of the Auger detector is taken from
Ref.~\cite{Capelle:98} (it is not assumed that a shower axis falls
certainly in the array). The threshold energy  $E_{th}$ is chosen
to be $10^{17}$ eV.

For comparison, a SM background is presented in the last row of
the Table~\ref{tab:cosmogenic flux}. This value is in agreement
with the number obtained recently for the same neutrino flux in
Ref.~\cite{Anchordoqui:04}.

\begin{table}[h!t]
\begin{center}
\caption{\small Yearly event rates for nearly horizontal neutrino
induced showers with $\theta_{zenith} > 70^{\circ}$ for the
cosmogenic neutrino flux from Ref.~\cite{Protheroe:96} for three
sets of the parameters. Number of events corresponds to one side
of the Auger ground array.}
\bigskip
  \begin{tabular}{||c||c|c|c||}
  \hline
   & $\Lambda_{\pi}$=2 TeV, $\mu$=0.10 & $\Lambda_{\pi}$=3 TeV, $\mu$=0.05 &
   $\Lambda_{\pi}$=3 TeV, $\mu$=0.10
  \\ \hline
  SM+grav & 0.81 & 0.66 & 0.43
  \\ \hline
  SM & \multicolumn{3}{c||}{0.24}
  \\
  \hline
  \end{tabular}
\label{tab:cosmogenic flux}
\end{center}
\end{table}

The cosmogenic neutrino flux is the most reliable one, since it
relies only on two assumptions: (i) the observed extremely high
energy cosmic rays contain protons, (ii) these cosmic rays are
primarily extragalactic in origin. Note, however, that the
cosmogenic neutrino flux may be significantly depleted, if a
substantial fraction of the cosmic ray primaries are heavy nuclei
rather than protons~\cite{Hooper:04}.

The cosmogenic neutrino flux not only represents a lower limit on
the flux of ultrahigh energy neutrinos, but it also can be used to
put an upper limit on the neutrino flux. In Ref.~\cite{Bahcall:01}
an upper limit (called WB bound) on a flux of neutrinos from
compact sources which are optically thin to $p \gamma$ and $pp$
interactions (such as active galactic nuclei) has been obtained.
The number of showers which can be registered by the Auger
detector for this case is shown in Table~\ref{tab:WB bound}. We
have chosen the same threshold energy $E_{th}$ = 10$^{17}$ eV and
put $E_{max} = 10^{21}$ eV.

\begin{table}[h!t]
\begin{center}
\caption{\small The same as in Table~\ref{tab:cosmogenic flux} but
for the Waxman-Bahcall neutrino flux~\cite{Bahcall:01}.}
\bigskip
  \begin{tabular}{||c||c|c|c||}
  \hline
   & $\Lambda_{\pi}$=2 TeV, $\mu$=0.10 & $\Lambda_{\pi}$=3 TeV, $\mu$=0.05 &
   $\Lambda_{\pi}$=3 TeV, $\mu$=0.10
  \\ \hline
  SM+grav & 1.03 & 0.78 & 0.49
  \\ \hline
  SM & \multicolumn{3}{c||}{0.28 }
  \\
  \hline
  \end{tabular}
\label{tab:WB bound}
\end{center}
\end{table}

Note, the so-called cascade upper limit on transparent neutrino
sources~\cite{Mannheim:01} (MPR bound) is 43 times higher that the
WB bound. It exploits the EGRET data on the diffuse gamma-ray
background~\cite{EGRET:98}.

The lower bound on the cosmogenic neutrino flux was also obtained
under assumption that the observed extremely high energy cosmic
rays below 10$^{20}$ eV are protons from uniformly distributed
extragalactic sources~\cite{Fodor:03}. It uses the fact that the
proton are accumulated around the energy $E_{_{GZK}} = 4 \cdot
10^{19}$ eV due to the GZK mechanism~\cite{Greisen:66}. The lower
cosmogenic neutrino spectrum is practically cut at $E_{\nu} \simeq
2 \cdot 10^{19}$ eV~\cite{Fodor:03}. Other recent estimates of the
cosmogenic neutrino fluxes can be found in
Refs.~\cite{Kalashev:02,Anchordoqui:04}.

\section{Conclusions and discussions}

In the present paper we have calculated the contribution from the
massive graviton modes to the eikonal in the RS model. The results
were applied to the neutrino-nucleon scattering at transplanckian
energies. Both differential and total cross sections are estimated
for the different sets of the parameters of the model. By using
differential cross sections, we have calculated the number of
quasi-horizontal neutrino induced air showers which can be
detected at the Auger Observatory per year. The estimates were
obtained for two fluxes of cosmogenic neutrinos.

\begin{figure}
\centering
\epsfig{figure=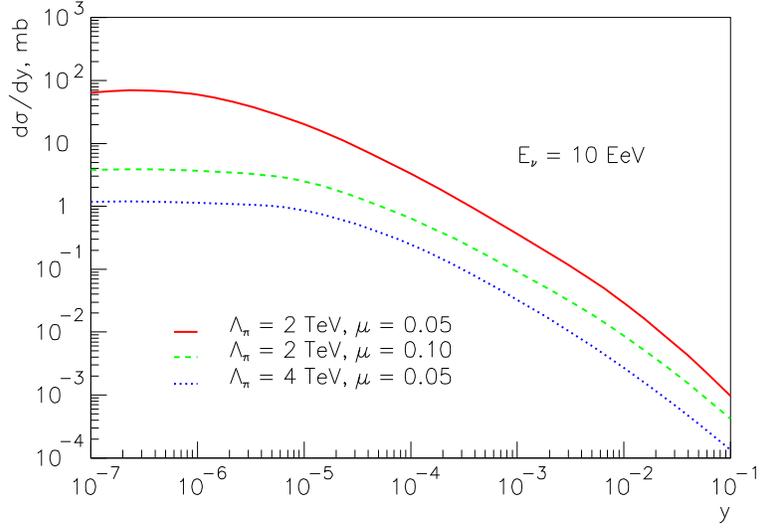,height=7cm,width=10cm}
\caption{The differential neutrino-proton cross section as a
function of $y$, the fraction of the neutrino energy deposited to
the proton.}
\label{fig:dsigma_10EeV}
\end{figure}

The differential cross section, $d\sigma(y)/dy$, where $y$ is the
fraction of the neutrino energy $E_{\nu}$ deposited to the shower,
can reach tens of mb at $y=0$, depending on energy
(Fig.~\ref{fig:dsigma_10EeV}). However, the differential cross
section exhibits a rapid fall-off in $y$, starting at some small
$y$. As a result, the gravitational cross section appears to be
approximately one order of magnitude larger than the SM cross
section at the same energy. To illustrate this statement, let us
fix the parameters of the RS model to be $\Lambda_{\pi} = 2$ TeV,
$\mu = 0.1$. Then we have $(d\sigma(y)/dy)|_{y=0} \simeq 4$ mb for
$E_{\nu} = 10^{10}$ GeV (see Fig.~\ref{fig:dsigma_10EeV}). As one
can see in Fig.~\ref{fig:dsigma_10EeV}, $d\sigma(y)/dy$ begins to
fall rapidly at $y > 10^{-5}$. The numerical calculations show
that $\sigma \simeq 4 \cdot 10^{-4}$ mb for this case (dashed
curve in Fig.~\ref{fig:sigma_y0_2}).

The energy of neutrino induced air shower, $E_{sh} = y E_{\nu}$,
is bounded from below by a threshold energy $E_{th}$. Thus, the
fraction $y$ should obey the inequality $y \geqslant
E_{th}/E_{max}$, where $E_{max}$ is a maximum energy in a neutrino
spectrum. For $E_{th} = 10^{8}$ GeV and $E_{max} = 10^{11(12)}$
GeV, we get $y \geqslant 10^{-3(4)}$. Thus, the air showers event
rate is defined by the region of $y$, in which neutrino-nucleon
cross section $d\sigma(y)/dy$ is significantly reduced in
comparison with its magnitude at $y=0$. Nevertheless, gravity
contribution to the event rate at the Auger detector is several
times larger than the SM background, as one can see from
Tables~\ref{tab:cosmogenic flux} and \ref{tab:WB bound}.

Recently, model independent bounds on the inelastic
neutrino-nucleon cross section derived from the AGASA~\cite{AGASA}
and RICE~\cite{RICE} search results on neutrino events were
obtained~\cite{Anchordoqui:04}. The bounds exploit the cosmogenic
neutrino fluxes from Refs.~\cite{Protheroe:96,Fodor:03}. However,
they were derived under an assumption that the \emph{total}
neutrino energy goes into shower energy, that is $y=1$. As we have
seen, it is not a case for the gravitational interactions
originated from
$t$-channel KK gravitons, which prefer $y \ll 1$.%
\footnote{In processes initiated by graviton $t$-channel exchanges
in large extra dimensions mean energy loss is also small, as was
pointed out in Ref.~\cite{Emparan:02}. On the contrary, in a
process of black hole production, the neutrino loses most of its
initial energy ($y \approx 1$).}
Generally, in order to extract an upper limit on $\sigma_{tot}$, a
dependence of $d\sigma(y)/dy$ on $y$ is needed. So, we conclude
that the bounds from Ref.~\cite{Anchordoqui:04} can not be
directly apply to the neutrino-nucleon cross sections derived in
our scheme.


\section*{Acknowledgments}

The author is indebted to V.A. Petrov for discussions and valuable
remarks.


\setcounter{equation}{0}
\renewcommand{\theequation}{A.\arabic{equation}}

\section*{Appendix}

In Appendix we calculate the dependence of the eikonal~\eqref{24}
on variable $u$~\eqref{22}. Let us define
\begin{equation}\label{A02}
I(u) = \int\limits_0^{\infty} \, dx \, J_0(u x) \, R(x),
\end{equation}
with
\begin{equation}\label{A04}
R(x) = \frac{I_2(x)}{I_1(x)}
\end{equation}
being the ratio of two modified Bessel functions.

It easily to see from \eqref{A02} that $I(u) \rightarrow u^{-1}$
at $u \rightarrow 0$, since $R(x) \rightarrow 1$ at $x \rightarrow
\infty$. The asymptotics of $I(u)$ at large $u$ is defined by the
behavior of the integrand at small $x$ which looks like
\begin{equation}\label{A06}
R(x) \simeq \frac{x}{4} - \frac{x^3}{96} + \frac{x^5}{1536} -
\frac{x^7}{23040} + \mathrm{O}(x^9).
\end{equation}

Let us now demonstrate that at $u \rightarrow \infty$ the function
$I(u)$~\eqref{A02} decreases faster that any fixed power of
$u^{-1}$. By using well-known relation~\cite{Erdelyi:II}
\begin{equation}\label{A08}
x^{\nu - 1}J_{\nu - 1}(u x) = \frac{1}{u} \, \Big( \frac{d}{x\,dx}
\Big) \big[ x^{\nu} J_{\nu}(u x)\big],
\end{equation}
and integrating \eqref{A02} by parts $k$ times, we obtain:
\begin{equation}\label{A09}
I(u) = \frac{1}{u^k} \int\limits_0^{\infty} \, dx \, J_k(u x)
F_k(x),
\end{equation}
where $J_k(z)$ is the Bessel function, and
\begin{equation}\label{A10}
F_k(x) = (-1)^k \, x^{k+1} \, \Big( \frac{d}{x\,dx} \Big)^k \left[
\frac{R(x)}{x} \right].
\end{equation}

The function $F_k(x)$ \eqref{A10} has the following properties: it
is proportional to $x^{k+1}$ at $x\rightarrow 0$, and it decreases
as $x^{-k}$ at $x \rightarrow \infty$. For any positive integer
$k$, it depends only on x and on the ratio $R(x)$~\eqref{A04} (but
not on $I_1(x)$ and $I_2(x)$ separately) due to the following
relations between modified Bessel functions~\cite{Erdelyi:II}:
\begin{eqnarray}\label{A11}
\frac{d}{dx} I_1(x) &=& I_2(x) +  \frac{1}{x} \, I_1(x),
\nonumber \\
\frac{d}{dx} I_2(x) &=& I_1(x) - \frac{2}{x} \, I_2(x).
\end{eqnarray}

For instance, for $k=1$ one has
\begin{equation}\label{A12}
F_1(x) = - 1 + \frac{4R(x)}{x}  + \big[ R(x) \big]^2
\end{equation}
This expression has asymptotics $x^2/48$ and $x^{-1}$ at small and
large $x$, respectively. For $k=2$ one gets
\begin{equation}\label{A14}
F_2(x) = \frac{1}{x} \Big\{ -6 \Big[ 1 - \frac{4R(x)}{x} \Big] -
2x \, R(x) + 12 \, \big[ R(x) \big]^2 + 2x \, \big[ R(x) \big]^3
\Big\}
\end{equation}
The asymptotics of $F_2(x)$ are $x^3/192$ and $3 \, x^{-2}$.

Since $k$ is an arbitrary positive integer, we conclude from
\eqref{A09} and \eqref{A10} that $\lim\limits_{u \rightarrow
\infty} u^a \,I(u) = 0$ for any $a > 0$.

The integral in \eqref{A09}, contrary to an original one
\eqref{A02}, converges rapidly at $x \rightarrow \infty$ for $k
\geqslant 2$, and could be used for numerical calculations.  It
cannot be calculated analytically. However, there exists an
expression which approximates our integral with a very high
accuracy:
\begin{equation}\label{A16}
\bar{I}(u) = \int\limits_0^{\infty} \, dx \, J_0(u x) \,
\bar{R}(x),
\end{equation}
with
\begin{equation}\label{A18}
\bar{R}(x) = \frac{\sqrt{3}}{2} \, \frac{x}{\sqrt{x^2 + 12}}.
\end{equation}
The function $\bar{R}(x)$ has the following expansion  at $x^2 <
12$ (compare with Eq.~\eqref{A06}):
\begin{equation}\label{A20}
\bar{R}(x) \simeq \frac{x}{4} - \frac{x^3}{96} + \frac{x^5}{1536}
- \frac{5 \, x^7}{110592} + \mathrm{O}(x^9).
\end{equation}
The integral \eqref{A16} is a table one~\cite{Prudnikov:II}:
\begin{equation}\label{A22}
\bar{I}(u) = \frac{\sqrt{3}}{2u} \, \exp(-2 \sqrt{3} \, u).
\end{equation}

\begin{figure}
\centering
\epsfig{figure=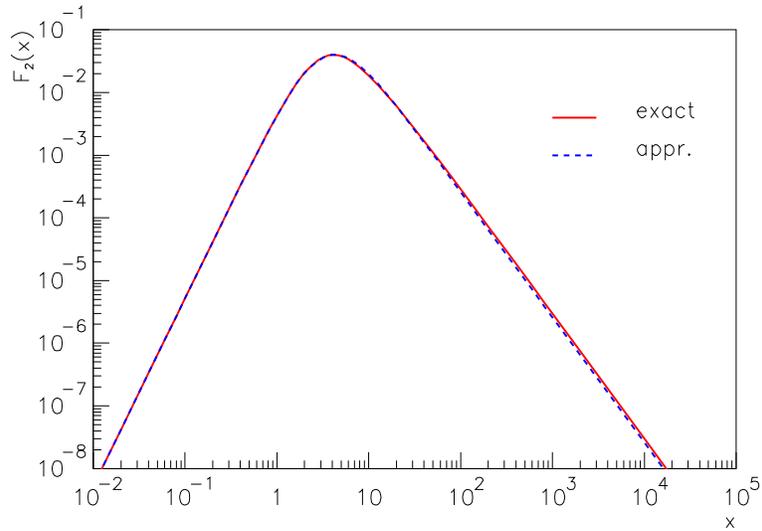,height=7cm,width=10cm}
\caption{The exact integrand  vs. approximate one (after the
integration of both integrals by parts twice). See Appendix for
details.}
\label{fig:integrand}
\end{figure}

\begin{figure}
\centering
\epsfig{figure=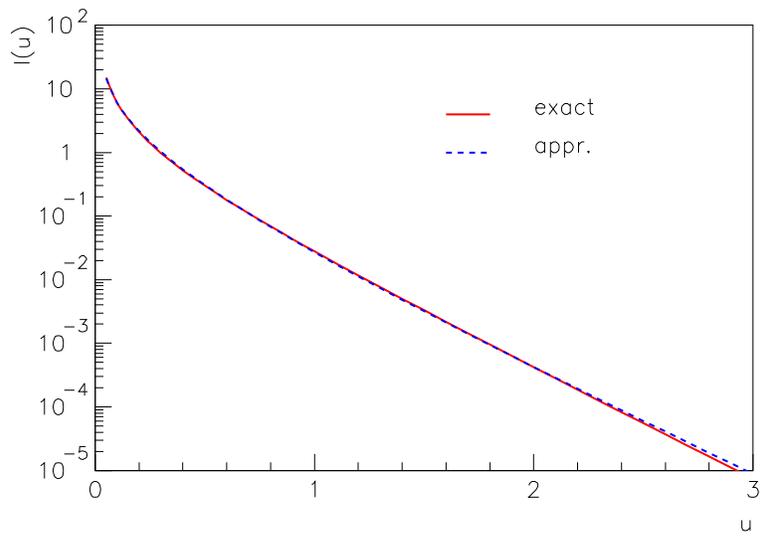,height=7cm,width=10cm}
\caption{The input integral vs. approximate one as a function
of the dimensionless variable $u$~\eqref{22}. See Appendix for
details.}
\label{fig:integral}
\end{figure}

We can integrate the RHS of \eqref{A16} twice by parts,
\begin{equation}\label{A24}
\bar{I}(u) = \frac{1}{u^2} \int\limits_0^{\infty} \, dx \, J_2(u
x) \, \bar{F}_2(x),
\end{equation}
and compare $\bar{F}_2(x) = (3\sqrt{3}/2)\,x^3/(x^2 + 12)^{5/2}$
with the corresponding function $F_2(x)$~\eqref{A14}. The result
of our calculations is presented in Fig.~\ref{fig:integrand}.
Formula \eqref{A22} gives practically the same dependence on
variable $u$ as a numerical integration of the exact expression by
using formula~\eqref{A09} (with $k=2$) does, see Fig.~\ref{fig:integral}.%
\footnote{Some disagreement between two curves at $u \gtrsim 2.5$
is not important, since $I(u)$ (and, consequently,
$\chi_{mass}(s,u)$) is strongly suppressed in this region. Note, a
region of very small $u$ also gives a negligible contribution to
the eikonal amplitude~\eqref{18}.}
Thus,  $I(u)$ exhibits an exponential fall-off (as we expected,
see above), and it becomes as small as $I(u) \simeq 0.01$ already
at $u=1.2$.

Taking all said above into account, we put $I(u) \rightarrow
\bar{I}(u)$, that results in the analytical expression for the
eikonal presented in the text~\eqref{25}.



\clearpage

\clearpage

\clearpage

\clearpage

\clearpage

\clearpage

\begin{thebibliography}{99}

\bibitem{Randall:99}
L. Randall and R. Sundrum, Phys. Rev. Lett.  {\bf 83} (1999) 3370.
\bibitem{Randall:99*}
L.~Randall and R.~Sundrum, Phys. Rev. Lett.  {\bf 83} (1999) 4690;
J. Lykken and L. Randall, JHEP {\bf 06} (2000) 014.
\bibitem{Davoudiasl:01}
H. Davoudiasl, J.L. Hewett and T.G. Rizzo, Phys. Rev. D {\bf 63}
(2001) 075004.
\bibitem{Davoudiasl:00}
H. Davoudiasl, J.L. Hewett and T.G. Rizzo, Phys. Rev. Lett. {\bf
84} (2000) 2080.
\bibitem{Allanach:00}
B.C. Allanach \emph{et al.}, JHEP {\bf 0009}  (2000) 019.
\bibitem{Goldberger:99}
W.D. Goldberger and M.B. Wise, Phys. Rev. Lett. {\bf 83}
(1999) 4922;
C. Cs\'{a}ki \emph{et al.}, Phys. Rev. D {\bf 62} (2000)  045015;
C. Cs\'{a}ki, M. Graesser and G.D. Kribs, \emph{ibid.} D {\bf 63}
(2001) 064020.
\bibitem{Watson}
G.N. Watson, \emph{A Treatise on the Theory of Bessel Functions}
(McMillan, 1922).
\bibitem{Han:99}
T. Han, J.D. Lykken and R.-J. Zhang, Phys. Rev. D {\bf 59} (1999)
105006.
\bibitem{Dvergsnes:03}
E. Dvergsnes, P. Osland and N. \"{O}zt\"{u}rk,  Phys. Rev. D {\bf
67} (2003)  074003.
\bibitem{Giudice:02}
G.F. Giudice, R. Rattazzi and J.D. Wells, Nucl. Phys. B {\bf 630}
(2002) 293.
\bibitem{Kisselev:03}
A.V. Kisselev, Eur. Phys. J. C {\bf 34} (2004) 513 .
\bibitem{Hooft:87}
G. 't Hooft, Phys. Lett. B {\bf 198} (1987) 61;
H. Verlinde and E. Verlinde, Nucl. Phys, B {\bf 371} (1992) 246;
M. Fabbrichesi \emph{et al.,} Nucl. Phys. B {\bf 419} (1994) 174.
\bibitem{Kisselev:04}
A.V. Kisselev and V.A. Petrov, Eur. Phys. J. C {\bf 36} (2004) 103;
Eur. Phys. J. C {\bf 37} (2004) 241.
\bibitem{Kisselev:05}
A.V. Kisselev and V.A. Petrov, in preparation.
\bibitem{Petrov:02}
V.A. Petrov, {\em Proceedings of the Int. Conf. Theor. Phys., TH
2002} (Paris, July 2002, Eds. D. Iagolnitzer, V. Rivasseau and J.
Zinn-Justin, Birkh\"{a}user Verlag, 2003). Supplement (2003) 253.
\bibitem{Alekhin:02}
S.I. Alekhin, Phys. Rev. D {\bf 68}, 014002 (2003).
\bibitem{Fly's Eye}
R.M. Baltrusaitis \emph{et al.}, Phys. Rev. D {\bf 31} (1985)
2192.
\bibitem{AGASA}
N. Inoue, \emph{Proc. 26th International Cosmic Ray Conference
(ICRC 1999)}, eds. D. Kieda, M. Salamon and B. Dingus, Salt Lake
city, Utah, 1999, v.~1, p.~361; S. Yoshida \emph{et al.},
\emph{Proc. 27th International Cosmic Ray Conference (ICRC 2001)},
Hamburg, Germany, 2001, v.~3, p.~1142
\bibitem{RICE}
I. Kravchenko \emph{et al.}, Astropart. Phys. {\bf 20} (2003) 195;
I. Kravchenko, arXiv: astro-ph/0306408.
\bibitem{Auger}
Pierre Auger Observatory, http://www.auger.org/
\bibitem{Sigl:98}
G. Sigl, Phys. Rev. D {\bf 57}, 3786 (1998).
\bibitem{Protheroe:96}
R.J. Protheroe and P.A. Johnson, Astropart. Phys. {\bf 4}, 253
(1996) [Erratum, {\em ibid.} {\bf 5}, 215 (1996)]; R.J. Protheroe,
Nucl. Proc. Suppl. {\bf 77}, 465 (1999).
\bibitem{Capelle:98}
K.S. Capelle, J.W. Cronin, G. Parente and E. Zas, Astrophys. Phys.
\bibitem{Anchordoqui:04}
L. Anchordoqui, Z. Fodor, S.D. Katz, A. Ringwald and H. Tu, arXiv:
hep-ph/0410136.
\bibitem{Hooper:04}
D. Hooper, A. Taylor and S. Sarkar, arXiv: astro-ph/0407618.
\bibitem{Bahcall:01}
J.N. Bahcall and E. Waxman, Phys. Rev. D {\bf 64} (2001) 023002.
\bibitem{Mannheim:01}
K. Mannheim, R.J. Protheroe and J.P. Rachen, Phys. Rev. D {\bf 63}
(2001) 023003.
\bibitem{EGRET:98}
P. Sreekumar \emph{et al.}, Astrophys. J. {\bf 494} (1998) 523.
\bibitem{Fodor:03}
Z. Fodor, S.D. Katz, A. Ringwald and H. Tu, JCAP {\bf 0311} (2003)
015;
A. Ringwald, arXiv: hep-ph/0409151.
\bibitem{Greisen:66}
K. Greisen, Phys. Rev. Lett. {\bf 16} (1966) 748;
G.T. Zatsepin and V.A. Kuzmin, JETP Lett. {\bf 4} (1966) 78.
\bibitem{Kalashev:02}
O.E. Kalashev, V.A. Kuzmin, D.V. Semikoz and G. Sigl, Phys.Rev.
D {\bf 66}, 063004 (2002); D.V. Semikoz and G. Sigl, JCAP {\bf
0404} (2004) 003.
\bibitem{Erdelyi:II}
A. Erdelyi, W. Magnus, F. Oberhettinger and F.C. Tricomi (eds.),
\emph{Higher Transcendental functions}, v.2 (Mc Graw-Hill Book H.
Company, 1955).
\bibitem{Prudnikov:II}
A.P. Prudnikov, Yu.A. Brychkov and O.I. Marichev, \emph{Integrals
and series}, v.2: \emph{Special functions}, Translated from
Russian (NY Gordon and Breach, 1986).
\bibitem{Emparan:02}
R. Emparan, M. Masip and R. Ratazzi, Phys. Rev. D {\bf 65} (2002)
064023.

\end{thebibliography}
\end{document}